\documentclass[a4paper, 10 pt, conference]{ieeeconf}
\IEEEoverridecommandlockouts
\usepackage{amsfonts,amsmath,amssymb} 

\usepackage{psfrag,color}
\usepackage{enumerate,cite,latexsym,graphicx}
\newtheorem{theorem}{Theorem}
\newtheorem{lemma}{Lemma}

\newtheorem{definition}{Definition}
\newtheorem{observation}{Observation}

\def\tr{\mathop{\rm Tr}\nolimits} 

\title{\LARGE \bf  Quantum Popov robust stability analysis of an optical cavity containing a saturated Kerr medium
}

\author{Ian R.~Petersen  %
\thanks{This work was supported by the
Australian Research Council (ARC) and Air Force Office of Scientific
Research (AFOSR). This material is based on research sponsored by the
Air Force Research Laboratory, under agreement number
FA2386-09-1-4089.  The U.S. Government is authorized to reproduce and
distribute reprints for Governmental purposes notwithstanding any
copyright notation thereon.
The views and conclusions contained herein are those of the authors
and should not be interpreted as necessarily representing the official
policies or endorsements, either expressed or implied, of the Air
Force Research Laboratory or the U.S. Government. }%
\thanks{Ian R. Petersen is with the School of  Engineering and Information Technology, 
        University of New South Wales at the Australian Defence Force Academy, Canberra ACT 2600, Australia.
         {\tt\small i.r.petersen@gmail.com} } 
}%

\begin{document}
\maketitle
\thispagestyle{empty}
\pagestyle{empty}

\begin{abstract}
This paper applies  results on the robust stability of nonlinear quantum systems to  a system consisting an optical cavity containing a saturated Kerr medium. The system is characterized by a Hamiltonian operator which contains a non-quadratic term involving a quartic function of the annihilation and creation operators. A saturated version of the Kerr nonlinearity leads to a sector bounded nonlinearity which enables a  quantum small gain theorem to be applied to this system in order to analyze its stability. Also, a non-quadratic version of a quantum Popov stability criterion is presented and applied to analyze the stability of this system. 
\end{abstract}

\section{Introduction} \label{sec:intro}
The use of Kerr media is commonly found in applications of nonlinear optics; e.g., see \cite{BOY08,NEW11}. A Kerr medium is characterized by a refractive index which increases with the intensity of light applied to the medium; e.g., see Section 9.1.1. of \cite{BR04}.  Within the area of quantum optics, a Kerr medium is often characterized by a
Hamiltonian operator which is a quartic function of the annihilation and creation operators; e.g., see Section 5.4 of \cite{WM08}. This leads to a nonlinear quantum stochastic differential equation which contains a cubic nonlinearity  \cite{WM08}. In this paper, we apply some recent and new quantum robust stability analysis tools to analyze the stability of an optical cavity containing a Kerr medium. Such as system has been proposed as a method of generating squeezed light; see Chapter 9 of \cite{BR04}. Squeezed light is an intrinsically quantum phenomenon which has potential applications in areas such as gravity wave detection, precision metrology and quantum computing \cite{BR04,WM10}. Also note that the quantum dynamics  obtained in the case of a microwave resonator containing a Josephson junction can be used to approximate the case of a Kerr medium in a cavity; e.g., see \cite{BOBBMDBVE12}. Such a system is related to the system analyzed in \cite{PET12Aa}. 

The first method we will apply to analyze the robust stability of the system under consideration is the quantum small gain result presented in \cite{PUJ1a}. This result gives a sufficient condition for the robust stability of uncertain nonlinear quantum systems in which the uncertainty is introduced by considering a non-quadratic perturbation to the system Hamiltonian operator. Such a non-quadratic perturbation leads to a nonlinear quantum stochastic differential equation describing the system; e.g., \cite{GJ09}. This nonlinearity is required to satisfy a certain sector bound condition. Related results to the results of \cite{PUJ1a} can be found in \cite{PUJ2,PUJ3a} and consider different classes of perturbations. Furthermore, the paper \cite{JPU1a} introduces a quantum version of the Popov stability criterion (e.g., see \cite{KHA02} for the classical Popov stability criterion), which allows for quadratic perturbations to the system Hamiltonian. That is, the paper \cite{JPU1a} considers uncertain quantum linear systems. In this paper we introduce a new version of the quantum Popov stability criterion which allows for non-quadratic perturbations to the system Hamiltonian and thus nonlinear uncertain quantum systems. As in \cite{PUJ1a},  the nonlinearity is required to satisfy a certain sector bound condition. This result is  applied to analyze the system consisting of an optical cavity containing a Kerr medium. 

For the quantum robust stability result introduced in \cite{PUJ1a} and in the new quantum Popov stability result introduced in this paper, the nominal quantum system is assumed to be a quantum linear system; e.g., see \cite{JNP1,NJP1,MaP3,MaP4,PET10Ba}. In addition, the nonlinearity is required to satisfy certain sector bound and smoothness conditions. However, for the standard quartic Hamiltonian model of a Kerr medium, the resulting cubic nonlinearity will not satisfy the sector bound conditions for any finite sector. We overcome this difficulty by noting that any practical implementation of a Kerr medium will not be precisely modelled by a quartic Hamiltonian but rather will suffer from some saturation effects; e.g., see \cite{BBBDS12}. This allows us to model the Kerr medium with a non-quadratic Hamiltonian such that the sector bound and smoothness conditions required in our quantum robust stability analysis results are satisfied. 

The remainder of the paper proceeds as follows. In Section
\ref{sec:systems}, we define the general class of nonlinear uncertain nonlinear quantum
systems under consideration. In this section, we also also recall the main result of \cite{PUJ1a} and present a new Popov type stability result for this class of nonlinear quantum systems.  In Section \ref{sec:example}, we analyze the system consisting of an  optical cavity containing a saturated Kerr nonlinearity using the two quantum robust stability analysis results presented.  In Section \ref{sec:conclusions},
we present some conclusions. The proofs of all of the main results are given in the Appendix.

\section{Robust Stability of Uncertain Nonlinear Quantum Systems} \label{sec:systems}
In this section, we describe the general class of quantum systems under consideration. 
As in the papers \cite{GJ09,JG10,PUJ1a,PUJ2,JPU1a},  we consider uncertain nonlinear open quantum systems defined by  parameters $(S,L,H)$ where $S$ is the scattering matrix which is typically chosen as the identity matrix, L is the coupling operator and $H$ is the system  Hamiltonian operator which is assumed to be of the form
\begin{equation}
\label{H1}
H = \frac{1}{2}\left[\begin{array}{cc}a^\dagger &
      a^T\end{array}\right]M
\left[\begin{array}{c}a \\ a^\#\end{array}\right]+f(z,z^*).
\end{equation}
Here $a$ is a vector of annihilation
operators on the underlying Hilbert space and $a^\#$ is the
corresponding vector of creation operators. Also,  $M \in \mathbb{C}^{2n\times 2n}$ is a Hermitian matrix of the
form
\begin{equation}
\label{Mform}
M= \left[\begin{array}{cc}M_1 & M_2\\
M_2^\# &     M_1^\#\end{array}\right]
\end{equation}
and $M_1 = M_1^\dagger$, $M_2 = M_2^T$.
In the case vectors of
operators, the notation  $^\dagger$ refers to the transpose of the vector of adjoint
operators and  in the case of matrices, this notation refers to the complex conjugate transpose of a matrix. In the case vectors of
operators, the notation $^\#$ refers to the vector of adjoint
operators and in the case of complex matrices, this notation refers to
the complex conjugate matrix. Also, the notation $^*$ denotes the adjoint of an
operator. The matrix $M$ is assumed to be known and defines the nominal quadratic part of the system Hamiltonian. 
  Furthermore, we assume the uncertain non-quadratic  part of the system Hamiltonian  $f(z,z^*)$ is defined by a formal power series of  the form
\begin{equation}
\label{H2nonquad}
f(z,z^*) = \sum_{k=0}^\infty\sum_{\ell=0}^\infty S_{k\ell}z^k(z^*)^\ell,
\end{equation}
which is assumed to converge in some suitable sense.
Here $S_{k\ell}=S_{\ell k}^*$,  and $z$ is a known scalar operator defined by
\begin{eqnarray}
\label{z}
z &=&  E_1a+E_2 a^\# \nonumber \\
&=& \left[\begin{array}{cc} E_1 & E_2 \end{array}\right]
\left[\begin{array}{c}a \\ a^\#\end{array}\right] = \tilde E 
\left[\begin{array}{c}a \\ a^\#\end{array}\right].
\end{eqnarray}
The term $f(z,z^*)$ is referred to as the perturbation Hamiltonian. It  is assumed to be unknown but is contained within a known set which will be defined below. Two different sets of perturbations will be considered depending on the robust stability condition which  is to be applied. 

 We assume the coupling operator $L$ is known and is of the form 
\begin{equation}
\label{L}
L = \left[\begin{array}{cc}N_1 & N_2 \end{array}\right]
\left[\begin{array}{c}a \\ a^\#\end{array}\right]
\end{equation}
where $N_1 \in \mathbb{C}^{m\times n}$ and $N_2 \in
\mathbb{C}^{m\times n}$. Also, we write
\[
\left[\begin{array}{c}L \\ L^\#\end{array}\right] = N
\left[\begin{array}{c}a \\ a^\#\end{array}\right] =
\left[\begin{array}{cc}N_1 & N_2\\
N_2^\# &     N_1^\#\end{array}\right]
\left[\begin{array}{c}a \\ a^\#\end{array}\right].
\]  

The annihilation and creation operators are assumed to satisfy the
canonical commutation relations:
\begin{eqnarray}
\label{CCR2}
\left[\left[\begin{array}{l}
      a\\a^\#\end{array}\right],\left[\begin{array}{l}
      a\\a^\#\end{array}\right]^\dagger\right]
&\stackrel{\Delta}{=}&\left[\begin{array}{l} a\\a^\#\end{array}\right]
\left[\begin{array}{l} a\\a^\#\end{array}\right]^\dagger
\nonumber \\
&&- \left(\left[\begin{array}{l} a\\a^\#\end{array}\right]^\#
\left[\begin{array}{l} a\\a^\#\end{array}\right]^T\right)^T\nonumber \\
&=& J
\end{eqnarray}
where $J = \left[\begin{array}{cc}I & 0\\
0 & -I\end{array}\right]$; e.g., see \cite{GGY08,GJN10,PET10Ba}.

To define the set of allowable perturbation Hamiltonians $f(\cdot)$, we first define the following formal partial derivatives:
\begin{equation}
\label{fdash}
\frac{\partial f(z,z^*)}{\partial z} \stackrel{\Delta}{=} \sum_{k=1}^\infty\sum_{\ell=0}^\infty k S_{k \ell} z^{k-1}(z^*)^\ell;
\end{equation}
\begin{equation}
\label{fddash}
\frac{\partial^2f(z,z^*)}{\partial z^2} \stackrel{\Delta}{=} \sum_{k=1}^\infty\sum_{\ell=0}^\infty k(k-1)S_{k\ell} z^{k-2}(z^*)^{\ell};
\end{equation}
\begin{equation}
\label{fddashm}
\frac{\partial^2f(z,z^*)}{\partial z\partial z ^*} \stackrel{\Delta}{=} \sum_{k=1}^\infty\sum_{\ell=1}^\infty k\ell S_{k\ell} z^{k-1}(z^*)^{\ell-1}.
\end{equation}
Then for given constants $\gamma > 0$, $\beta > 0$, $\delta_1\geq 0$, $\delta_2\geq 0$, $\delta_3\geq 0$, we consider the sector bound conditions
\begin{equation}
\label{sector1}
\frac{\partial f(z,z^*)}{\partial z}^*\frac{\partial f(z,z^*)}{\partial z}  \leq \frac{1}{\gamma^2}z z^* + \delta_1;
\end{equation}
\begin{eqnarray}
\label{sector2}
\lefteqn{\left(\frac{\partial f(z,z^*)}{\partial z}
-\frac{1}{\gamma}z^*\right)^*\left(\frac{\partial f(z,z^*)}{\partial z}-\frac{1}{\gamma}z^*\right) } \nonumber \\
 &\leq& \frac{1}{\gamma^2}z z^* + \delta_1; \hspace{5cm}
\end{eqnarray}
and the smoothness conditions
\begin{eqnarray}
\label{smooth1}
\frac{\partial^2f(z,z^*)}{\partial z^2}^*\frac{\partial^2f(z,z^*)}{\partial z^2} \leq  \delta_2;\\
\label{smooth2}
\frac{\partial^2f(z,z^*)}{\partial z\partial z ^*}^*
\frac{\partial^2f(z,z^*)}{\partial z\partial z ^*}\leq  \delta_3.
\end{eqnarray}
Also, we consider the 
 following upper and lower bounds on the perturbation Hamiltonian
\begin{equation}
\label{upper_lower}
0 \leq f(z,z^*) \leq \beta z z^*.
\end{equation}

Then we define two possible sets of perturbation Hamiltonians $\mathcal{W}_1$ and  $\mathcal{W}_2$ as follows:
\begin{equation}
\label{W1}
\mathcal{W}_1 = \left\{\begin{array}{l}f(\cdot) \mbox{ of the form
      (\ref{H2nonquad}) such that 
} \\
\mbox{ conditions (\ref{sector1}) and (\ref{smooth1}) are satisfied}\end{array}\right\};
\end{equation}
\begin{equation}
\label{W2}
\mathcal{W}_2 = \left\{\begin{array}{l}f(\cdot) \mbox{ of the form
      (\ref{H2nonquad}) such that 
} \\
\mbox{ conditions (\ref{sector2}), (\ref{smooth1}), (\ref{smooth2}) and (\ref{upper_lower}) }\\
\mbox{ are satisfied}\end{array}\right\}.
\end{equation}

As in \cite{PUJ1a,PUJ2,JPU1a}, we consider a notion of robust mean square stability. 
\begin{definition}
\label{D1}
An uncertain open quantum system defined by  $(S,L,H)$ where $H$ of the form (\ref{H1}), $f(\cdot) \in \mathcal{W}$, and $L$  of the form (\ref{L}) is said to be {\em robustly mean square stable} if there exist constants $c_1 > 0$, $c_2 > 0$ and $c_3 \geq 0$ such that for any $f(\cdot) \in \mathcal{W}$,  
\begin{eqnarray}
\label{ms_stable0}
\lefteqn{\left< \left[\begin{array}{c}a(t) \\ a^\#(t)\end{array}\right]^\dagger \left[\begin{array}{c}a(t) \\ a^\#(t)\end{array}\right] \right>}\nonumber \\
&\leq& c_1e^{-c_2t}\left< \left[\begin{array}{c}a \\ a^\#\end{array}\right]^\dagger \left[\begin{array}{c}a \\ a^\#\end{array}\right] \right>
+ c_3~~\forall t \geq 0.
\end{eqnarray}
Here $\left[\begin{array}{c}a(t) \\ a^\#(t)\end{array}\right]$ denotes the Heisenberg evolution of the vector of operators $\left[\begin{array}{c}a \\ a^\#\end{array}\right]$; e.g., see \cite{JG10}.
\end{definition}

  The following small gain condition
is sufficient for the robust mean square stability
of the nonlinear quantum system under consideration when $f(\cdot) \in \mathcal{W}_1$: 
\begin{enumerate}
\item
The matrix 
\begin{equation}
\label{Hurwitz1}
F = -iJM-\frac{1}{2}JN^\dagger J N\mbox{ is Hurwitz;}
\end{equation}
\item
The transfer function 
\begin{equation}
\label{Gs}
G(s) =  2i \tilde E^\# \Sigma \left(sI-F\right)^{-1} \Sigma J \tilde E^T
\end{equation}
satisfies the $H^\infty$ norm bound
\begin{equation}
\label{Hinfbound1}
\left\|G(s)\right\|_\infty < \gamma.
\end{equation}
Here,
\begin{eqnarray*}
\Sigma &=& \left[\begin{array}{cc} 0 & I\\
I &0 \end{array}\right].
\end{eqnarray*}
\end{enumerate}
This result is given in  the following theorem which is presented in \cite{PUJ1a}.
\begin{theorem}
\label{T1}
Consider an uncertain open nonlinear quantum system defined by $(S,L,H)$  such that
$H$ is of the form (\ref{H1}), $L$ is of the
form (\ref{L}) and $f(\cdot) \in \mathcal{W}_1$. Furthermore, assume that
the strict bounded real condition  (\ref{Hurwitz1}), (\ref{Hinfbound1})
is satisfied. Then the
uncertain quantum system is robustly mean square stable. 
\end{theorem}

In the next section, we will apply this theorem to analyze the robust
stability of a nonlinear quantum system corresponding to an optical cavity containing a Kerr medium.  
We  also consider  a new sufficient condition for robust mean square stability   when $f(\cdot) \in \mathcal{W}_2$, which is a nonlinear quantum version of the Popov stability criterion. This new condition is  the existence of a constant $\theta \geq 0$, such that the matrix $F$ defined in (\ref{Hurwitz1}) 
is Hurwitz
and the transfer function $G(s)$ defined in (\ref{Gs}) 
satisfies the strict positive real condition
\begin{eqnarray}
\label{SPR}
&&\gamma + (1+\theta i \omega)G(i\omega) 
+ (1-\theta i \omega)G(i\omega)^* > 0
\end{eqnarray}
for all $\omega \in [-\infty,\infty]$. 
This result is given in  the following theorem.
\begin{theorem}
\label{T2}
Consider an uncertain open nonlinear quantum system defined by $(S,L,H)$  such that
$H$  is of the form (\ref{H1}), $L$ is of the
form (\ref{L}) and $f(\cdot) \in \mathcal{W}_2$. Furthermore, assume that there exists a constant $\theta \geq 0$ 
such that the  matrix $F$ defined in (\ref{Hurwitz1}) 
is Hurwitz and the frequency domain condition (\ref{SPR})
is satisfied. Then the
uncertain quantum system is robustly mean square stable. 
\end{theorem}
In order to prove this theorem, we require the following definitions and lemmas.

\begin{lemma}[See Lemma 3.4 of \cite{JG10}.]
\label{L0}
Consider an open quantum system defined by $(S,L,H)$ and suppose there exists a non-negative self-adjoint operator $V$ on the underlying Hilbert space such that
\begin{equation}
\label{lyap_ineq}
-\imath[V,H] + \frac{1}{2}L^\dagger[V,L]+\frac{1}{2}[L^\dagger,V]L + cV \leq \lambda
\end{equation}
where $c > 0$ and $\lambda$ are real numbers. 
Then for any plant state, we have
\[
\left<V(t)\right> \leq e^{-ct}\left<V\right> + \frac{\lambda}{c},~~\forall t \geq 0.
\]
\end{lemma}
In the above lemma, $[\cdot,\cdot]$ denotes the commutator between two operators. In the case of a commutator between a scalar operator and a vector of operators, this notation denotes the corresponding vector of commutator operators. Also, $V(t)$ denotes the Heisenberg evolution of the operator $V$ and $\left<\cdot\right>$ denotes quantum expectation; e.g., see \cite{JG10}.

We will consider  ``Lyapunov'' operators  $V$ of the form 
\begin{equation}
\label{quadV}
V = \left[\begin{array}{cc}a^\dagger &
      a^T\end{array}\right]P
\left[\begin{array}{c}a \\ a^\#\end{array}\right]+\theta f(z,z^*).
\end{equation}
where $P \in \mathbb{C}^{2n\times 2n}$ is a positive-definite Hermitian matrix of the
form
\begin{equation}
\label{Pform}
P= \left[\begin{array}{cc}P_1 & P_2\\
P_2^\# &     P_1^\#\end{array}\right]
\end{equation}
and $\theta \geq 0$.
 Hence, we consider a set of  non-negative self-adjoint operators
$\mathcal{P}$ defined as
\begin{equation}
\label{P1}
\mathcal{P} = \left\{\begin{array}{l}V \mbox{ of the form
      (\ref{quadV}) such that $P > 0$ is a 
} \\
\mbox{  Hermitian matrix of the form (\ref{Pform})}\end{array}\right\}.
\end{equation}

\begin{lemma}
\label{L4}
Given any positive definite matrix $P$ of the form (\ref{Pform}), then
\begin{eqnarray}
\label{mui}
\mu &=& \left[z,[z,\left[\begin{array}{cc}a^\dagger &
      a^T\end{array}\right]P
\left[\begin{array}{c}a \\ a^\#\end{array}\right]]\right]\nonumber \\
& =& \left[z^*,[z^*,\left[\begin{array}{cc}a^\dagger &
      a^T\end{array}\right]P
\left[\begin{array}{c}a \\ a^\#\end{array}\right]]\right]^* \nonumber \\
&=& 
-\tilde E \Sigma JPJ\tilde E^T,
\end{eqnarray}
which is a constant.  
\end{lemma}
{\em Proof:}
The proof of this result follows via a straightforward but tedious
calculation using (\ref{CCR2}). \hfill $\Box$

\begin{lemma}
\label{L5}
With the variable $z$ defined as in (\ref{z}) and $L$ defined as in (\ref{L}), then
\[
[z,L]  = \left[\tilde E \left[\begin{array}{c}a \\ a^\#\end{array}\right], \tilde N\left[\begin{array}{c}a \\ a^\#\end{array}\right]\right] = \tilde N \Sigma J \tilde E^T
\]
which is a constant vector. Here, 
\[
\Sigma = \left[\begin{array}{cc}0 & I\\I & 0\end{array}\right].
\]
Similarly
\[
[z^*,L]  = \left[\tilde E^\# \Sigma \left[\begin{array}{c}a \\ a^\#\end{array}\right], \tilde N\left[\begin{array}{c}a \\ a^\#\end{array}\right]\right] = -\tilde N  J \tilde E^\dagger
\]
which is a constant vector.

In addition
\begin{eqnarray*}
\left[[L^\dagger,z]L,z\right] &=&  \left[[z^*,L]^\dagger L,z\right] \\
&=& \left[[z^*,L]^\dagger \tilde N
\left[\begin{array}{c}a \\ a^\#\end{array}\right],
\tilde E \left[\begin{array}{c}a \\ a^\#\end{array}\right]\right]\\
&=& \tilde E \Sigma J \tilde N^T [z^*,L]^\# = -\tilde E \Sigma J \tilde N^T \tilde N^\# J \tilde E^T
\end{eqnarray*}
and
\begin{eqnarray*}
\left[z^*,[L^\dagger,z]L\right] &=& \left[z^*,[z^*,L]^\dagger L \right] \\
&=& \left[\tilde E^\#\Sigma \left[\begin{array}{c}a \\ a^\#\end{array}\right],
[z^*,L]^\dagger \tilde N
\left[\begin{array}{c}a \\ a^\#\end{array}\right]
\right]\\
&=& [z^*,L]^\dagger \tilde N \Sigma J \Sigma \tilde E^\dagger = \tilde E  J \tilde N^\dagger  \tilde N J \tilde E^\dagger
\end{eqnarray*}
which are constants. 
\end{lemma}
{\em Proof:}
The proofs of these equations  follows via  straightforward but tedious
calculations using (\ref{CCR2}). \hfill $\Box$

\begin{lemma}
\label{LB}
Given any  Hermitian matrix $\tilde P$ of the form (\ref{Pform}),  then the Hermitian operator
\[
\tilde V = \left[\begin{array}{cc}a^\dagger &
      a^T\end{array}\right]\tilde P 
\left[\begin{array}{c}a \\ a^\#\end{array}\right]
\]
satisfies
\begin{eqnarray}
\label{comm_condition}
[\tilde V,f(z,z^*)] &=& [\tilde V,z]w_1-w_1^*[z^*,\tilde V]
\nonumber \\&&
+\frac{1}{2}\left[z,[\tilde V,z]\right]w_2
-\frac{1}{2}w_2^*\left[z,[\tilde V,z]\right]^*,\nonumber \\
\end{eqnarray} 
for all $f(z,z^*) \in \mathcal{W}_2$
where
\begin{eqnarray}
\label{w1w2}
w_1 &=& \frac{\partial f(z,z^*)}{\partial z},~w_2=  \frac{\partial^2f(z,z^*)}{\partial z^2}
\end{eqnarray}
and the constant $\mu$ is defined as in (\ref{mui}). 
\end{lemma}

\noindent
{\em Proof:}
First,  given any  $k \geq 1$ note that 
\begin{eqnarray}
\label{Vzetak}
 \tilde Vz  &=& [\tilde V,z]+ z \tilde V;\nonumber \\
\vdots && \nonumber \\
\tilde Vz^k &=& \sum_{n=1}^k z^{n-1}[\tilde V,z] z^{k-n}+z^k \tilde V.
\end{eqnarray}
Also for any $n \geq 1$,
\begin{eqnarray}
\label{Vzetak1}
z [\tilde V,z] &=& [\tilde V,z]z + \left[z,[\tilde V,z]\right]; \nonumber \\
\vdots && \nonumber \\
 z^{n-1} [\tilde V,z]&=&[\tilde V,z]z^{n-1} \nonumber \\
&&+ (n-1)z^{n-2}\left[z,[\tilde V,z]\right].
\end{eqnarray}
Therefore using (\ref{Vzetak}) and (\ref{Vzetak1}), it follows that
\begin{eqnarray*}
\tilde Vz^k &=&\sum_{n=1}^k [\tilde V,z] z^{n-1}z^{k-n}  \nonumber \\
&&+ (n-1)z^{n-2}z^{k-n}\left[z,[\tilde V,z]\right]+z^k \tilde V\nonumber \\
&=& \sum_{n=1}^k [\tilde V,z] z^{k-1}+ (n-1)z^{k-2}\left[z,[\tilde V,z]\right] \nonumber \\&&
 +z^k \tilde V\nonumber \\
&=&k[\tilde V,z] z^{k-1}+\frac{k(k-1)}{2}z^{k-2}\left[z,[\tilde V,z]\right]\nonumber \\&&
+z^k \tilde V
\end{eqnarray*}
which holds for any $k \geq 0$. 
Similarly
\begin{eqnarray*}
(z^*)^k \tilde V &=& k(z^*)^{k-1}[z^*,\tilde V]
\nonumber \\
&&+\frac{k(k-1)}{2}\left[z,[\tilde V,z]\right]^*(z^*)^{k-2}\nonumber \\&&
+\tilde V(z^*)^k.
\end{eqnarray*}

Now given any  $k \geq 0$ and $\ell \geq 0$, let $H_{k\ell}= z^k(z^*)^\ell$ and we have
\begin{eqnarray}
\label{VHkl}
[\tilde V,H_{k\ell}] &=& k[\tilde V,z] z^{k-1}(z^*)^\ell \nonumber \\&&
+\frac{k(k-1)}{2}\left[z,[\tilde V,z]\right]z^{k-2}(z^*)^\ell\nonumber \\&&
+z^k \tilde V(z^*)^\ell\nonumber \\
&& -kz^\ell(z^*)^{k-1}[z^*,\tilde V]\nonumber \\&&
-\frac{k(k-1)}{2}\left[z,[\tilde V,z]\right]^*z^\ell(z^*)^{k-2}\nonumber \\&&
-z^\ell \tilde V(z^*)^k\nonumber \\
&=& k[\tilde V,z] z^{k-1}(z^*)^\ell - kz^\ell(z^*)^{k-1}[z^*,\tilde V]\nonumber \\
&&+\frac{k(k-1)}{2}\left[z,[\tilde V,z]\right]z^{k-2}(z^*)^\ell\nonumber \\
&&-\frac{k(k-1)}{2}\left[z,[\tilde V,z]\right]^*z^\ell(z^*)^{k-2}.\nonumber \\
\end{eqnarray}
Therefore, given any $f(z,z^*) \in \mathcal{W}_2$,
\begin{eqnarray}
\label{VH2}
[\tilde V,f(z,z^*)] &=& \sum_{k=0}^\infty \sum_{\ell=0}^\infty S_{k\ell} [\tilde V,H_{k\ell}] \nonumber \\
&=& [\tilde V,z]\frac{\partial f(z,z^*)}{\partial z}-\frac{\partial f(z,z^*)}{\partial z}^*[z^*,\tilde V]\nonumber \\
&&+ \frac{1}{2}\left[z,[\tilde V,z]\right] \frac{\partial^2f(z,z^*)}{\partial z^2}\nonumber \\&&
-\frac{1}{2}\left[z,[\tilde V,z]\right]^* \frac{\partial^2f(z,z^*)}{\partial z^2}^*.
\end{eqnarray}
Hence using (\ref{w1w2}), the condition (\ref{comm_condition}) is satisfied. \hfill $\Box$

\begin{lemma}
\label{LB1}
Given any  $f(z,z^*) \in \mathcal{W}_2$ and $L$ defined as in (\ref{L}), then
\begin{eqnarray}
\label{comm_condition1}
\lefteqn{\frac{1}{2}L^\dagger[f(z,z^*),L] + \frac{1}{2}[L^\dagger,f(z,z^*)]}\nonumber \\
&= &\frac{1}{2}\left(L^\dagger[z,L] + [L^\dagger,z]L\right)w_1\nonumber \\
&&+\frac{1}{2}w_1^* \left(L^\dagger[z^*,L] + [L^\dagger,z^*]L\right) \nonumber \\
&& -\frac{1}{2}\left[[L^\dagger,z]L,z\right]w_2\nonumber \\
&&-\frac{1}{2}w_2^*\left[[L^\dagger,z]L,z\right]^*\nonumber \\
&& +\frac{1}{2}\left[z^*,[L^\dagger,z]L\right]w_3\nonumber \\
&&+\frac{1}{2}w_3^*\left[z^*,[L^\dagger,z]L\right]^*
\end{eqnarray}
where
\begin{eqnarray}
\label{w1w2w3}
w_1 &=& \frac{\partial f(z,z^*)}{\partial z},~w_2=  \frac{\partial^2f(z,z^*)}{\partial z^2},~
w_3=  \frac{\partial^2 f(z,z^*)}{\partial z\partial z^*}.\nonumber \\
\end{eqnarray}
\end{lemma}

\noindent
{\em Proof:}
In a similar fashion to the proof of Lemma \ref{LB}, we write
\begin{eqnarray}
\label{VzetakL}
 Lz  &=& -[z,L]+ z L;\nonumber \\
\vdots && \nonumber \\
Lz^k &=& -\sum_{n=1}^k [z,L] z^{k-1}+z^k L\nonumber \\
&=& -[z,L] kz^{k-1}+z^k L.
\end{eqnarray}
Similarly
\begin{eqnarray*}
z^*L  &=& [z^*,L]+  Lz^*;\nonumber \\
\vdots && \nonumber \\
(z^*)^\ell L &=& \sum_{n=1}^k [z^*,L] (z^*)^{\ell-1}+L(z^*)^\ell\nonumber \\
&=& [z^*,L] \ell(z^*)^{\ell-1}+L(z^*)^\ell.
\end{eqnarray*}

Now given any $k \geq 0$ and $\ell \geq 0$, let $H_{k\ell}= z^k(z^*)^\ell$ and we have
\begin{eqnarray}
\label{VHklL}
[L,H_{k\ell}] &=& Lz^k(z^*)^\ell -z^k(z^*)^\ell L\nonumber \\
&=&-[z,L]kz^{k-1}(z^*)^\ell
+z^k L(z^*)^\ell\nonumber \\
&& -[z^*,L]\ell z^k(z^*)^{\ell-1}
-z^k L(z^*)^\ell\nonumber \\
&=& -[z,L]kz^{k-1}(z^*)^\ell
-[z^*,L]\ell z^k(z^*)^{\ell-1}.\nonumber \\ 
\end{eqnarray}
Therefore, given any $f(z,z^*) \in \mathcal{W}_2$,
\begin{eqnarray}
\label{VH2L}
[L,f(z,z^*)] &=& \sum_{k=0}^\infty \sum_{\ell=0}^\infty S_{k\ell} [L,H_{k\ell}] \nonumber \\
&=& -[z,L]\frac{\partial f(z,z^*)}{\partial z}-[z^*,L]\frac{\partial f(z,z^*)}{\partial z}^*. \nonumber \\
\end{eqnarray}

We now let $\rho = [L^\dagger,z]L$, which is a scalar operator and consider $[\rho,\frac{\partial f(z,z^*)}{\partial z}]$. Now Lemma \ref{L5} implies that $\left[[L^\dagger,z]L,z\right] = [\rho,z]$ a constant, and $\left[z^*,[L^\dagger,z]L\right] = [z^*,\rho]$, a constant. Then,
\begin{eqnarray}
\label{rhozeta}
 \rho z  &=& \left[[L^\dagger,z]L,z\right]+ z \rho;\nonumber \\
\vdots && \nonumber \\
\rho z^{k-1} &=& \sum_{n=1}^{k-1} \left[[L^\dagger,z]L,z\right] z^{k-2}+z^{k-1} \rho\nonumber \\
&=& \left[[L^\dagger,z]L,z\right] (k-1)z^{k-2}+z^{k-1} \rho.
\end{eqnarray}
Similarly
\begin{eqnarray*}
z^*\rho  &=& \left[z^*,[L^\dagger,z]L\right]+  \rho z^*;\nonumber \\
\vdots && \nonumber \\
(z^*)^\ell \rho &=& \sum_{n=1}^k \left[z^*,[L^\dagger,z]L\right] (z^*)^{\ell-1}+\rho(z^*)^\ell\nonumber \\
&=& \left[z^*,[L^\dagger,z]L\right] \ell(z^*)^{\ell-1}+\rho(z^*)^\ell.
\end{eqnarray*}

Now given any $f(z,z^*) \in \mathcal{W}_2$, $k \geq 0$, $\ell \geq 0$,  we have
\begin{eqnarray}
\label{rhoell}
[\rho,z^{k-1}(z^*)^\ell] &=& \rho z^{k-1}(z^*)^\ell -z^{k-1}(z^*)^\ell \rho\nonumber\\
&=&\left[[L^\dagger,z]L,z\right](k-1)z^{k-2}(z^*)^\ell \nonumber \\ &&
+z^{k-1} \rho(z^*)^\ell\nonumber \\
&& -\left[z^*,[L^\dagger,z]L\right]\ell z^{k-1}(z^*)^{\ell-1}\nonumber \\ &&
-z^{k-1} \rho(z^*)^\ell\nonumber \\
&=& \left[[L^\dagger,z]L,z\right](k-1)z^{k-2}(z^*)^\ell\nonumber \\ &&
-\left[z^*,[L^\dagger,z]L\right]\ell z^{k-1}(z^*)^{\ell-1}.\nonumber \\
\end{eqnarray}
Therefore,
\begin{eqnarray}
\label{rhow1}
\lefteqn{[L^\dagger,z]L\frac{\partial f(z,z^*)}{\partial z}-\frac{\partial f(z,z^*)}{\partial z}[L^\dagger,z]L}\nonumber \\
 &=& 
[\rho,\frac{\partial f(z,z^*)}{\partial z}] \nonumber \\
&=&\sum_{k=0}^\infty \sum_{\ell=0}^\infty kS_{k\ell} [\rho,z^{k-1}(z^*)^\ell] \nonumber \\
&=&  \left[[L^\dagger,z]L,z\right]\sum_{k=0}^\infty \sum_{\ell=0}^\infty k(k-1)S_{k\ell}z^{k-2}(z^*)^\ell
\nonumber \\&&
-\left[z^*,[L^\dagger,z]L\right]\sum_{k=0}^\infty \sum_{\ell=0}^\infty k\ell S_{k\ell} 
z^{k-1}(z^*)^{\ell-1}\nonumber \\
&=& \left[[L^\dagger,z]L,z\right]\frac{\partial^2 f(z,z^*)}{\partial z^2}
-\left[z^*,[L^\dagger,z]L\right]\frac{\partial^2 f(z,z^*)}{\partial z\partial z^*}.
\nonumber \\
\end{eqnarray}

Similarly
\begin{eqnarray}
\label{rhow1t}
\lefteqn{\frac{\partial f(z,z^*)}{\partial z}^*L^\dagger[z^*,L]- L^\dagger[z^*,L]\frac{\partial f(z,z^*)}{\partial z}^*}\nonumber \\
&=& \frac{\partial^2 f(z,z^*)}{\partial z^2}^*\left[[L^\dagger,z]L,z\right]^* \nonumber \\
&&-\frac{\partial^2 f(z,z^*)}{\partial z\partial z^*}^*\left[z^*,[L^\dagger,z]L\right]^*.
\nonumber \\
\end{eqnarray}

Now using (\ref{VH2L}), 
 it follows that
\begin{eqnarray*}
\lefteqn{\frac{1}{2}L^\dagger[f(z,z^*),L] + \frac{1}{2}[L^\dagger,f(z,z^*)]}\nonumber \\
 &=& \frac{1}{2}L^\dagger[f(z,z^*),L] + \frac{1}{2}[L^\dagger,f(z,z^*)]L\nonumber \\
&=& \frac{1}{2}L^\dagger\left(\begin{array}{c} [z,L]\frac{\partial f(z,z^*)}{\partial z}\nonumber\\
+[z^*,L]\frac{\partial f(z,z^*)}{\partial z}^*  \end{array}\right)\nonumber\\
&&\frac{1}{2}\left(\begin{array}{c}\frac{\partial f(z,z^*)}{\partial z}^* [z,L]^\dagger \nonumber\\
+\frac{\partial f(z,z^*)}{\partial z}[z^*,L]^\dagger \end{array}\right)L\nonumber\\
&=& \frac{1}{2}L^\dagger [z,L]\frac{\partial f(z,z^*)}{\partial z} \nonumber \\
&&+\frac{1}{2}L^\dagger[z^*,L]\frac{\partial f(z,z^*)}{\partial z}^* \nonumber \\
&& +\frac{1}{2}\frac{\partial f(z,z^*)}{\partial z}^* [z,L]^\dagger L \nonumber \\
&& +\frac{1}{2}\frac{\partial f(z,z^*)}{\partial z}[z^*,L]^\dagger L.
\end{eqnarray*}
Hence using (\ref{rhow1}) and (\ref{rhow1t}), we have
\begin{eqnarray}
\label{MLH2}
\lefteqn{\frac{1}{2}L^\dagger[f(z,z^*),L] + \frac{1}{2}[L^\dagger,f(z,z^*)]}\nonumber \\
&=& \frac{1}{2}L^\dagger [z,L]\frac{\partial f(z,z^*)}{\partial z} \nonumber \\
&&+\frac{1}{2}\frac{\partial f(z,z^*)}{\partial z}^*L^\dagger[z^*,L]\nonumber \\
&&- \frac{1}{2}\frac{\partial^2 f(z,z^*)}{\partial z^2}^*\left[[L^\dagger,z]L,z\right]^* \nonumber \\
&&+\frac{1}{2}\frac{\partial^2 f(z,z^*)}{\partial z\partial z^*}^*\left[z^*,[L^\dagger,z]L\right]^*\nonumber \\
&& +\frac{1}{2}\frac{\partial f(z,z^*)}{\partial z}^* [z,L]^\dagger L \nonumber \\
&&+\frac{1}{2}[L^\dagger,z]L\frac{\partial f(z,z^*)}{\partial z}\nonumber \\
&& -\frac{1}{2}\left[[L^\dagger,z]L,z\right]\frac{\partial^2 f(z,z^*)}{\partial z^2} \nonumber \\
&&+\frac{1}{2}\left[z^*,[L^\dagger,z]L\right]\frac{\partial^2 f(z,z^*)}{\partial z\partial z^*} \nonumber \\
&=& \frac{1}{2}\left(L^\dagger [z,L]+[L^\dagger,z]L \right)\frac{\partial f(z,z^*)}{\partial z} \nonumber \\
&&+\frac{1}{2}\frac{\partial f(z,z^*)}{\partial z}^*\left(L^\dagger[z^*,L]+[L^\dagger,z^*] L\right)\nonumber \\
&& -\frac{1}{2}\left[[L^\dagger,z]L,z\right]\frac{\partial^2 f(z,z^*)}{\partial z^2} \nonumber \\
&&- \frac{1}{2}\frac{\partial^2 f(z,z^*)}{\partial z^2}^*\left[[L^\dagger,z]L,z\right]^* \nonumber \\
&&+\frac{1}{2}\left[z^*,[L^\dagger,z]L\right]\frac{\partial^2 f(z,z^*)}{\partial z\partial z^*} \nonumber \\
&&+\frac{1}{2}\frac{\partial^2 f(z,z^*)}{\partial z\partial z^*}^*\left[z^*,[L^\dagger,z]L\right]^*.
\end{eqnarray}
It follows using (\ref{w1w2w3}) that the condition (\ref{comm_condition1}) is satisfied.
\hfill $\Box$

\begin{lemma}
\label{L2}
Given a positive definite matrix $P$ of the form (\ref{Pform}), a Hermitian matrix $M$ of the form (\ref{Mform}), and $L$ defined as in (\ref{L}), then
\begin{eqnarray*}
\lefteqn{ \left[\left[\begin{array}{cc}a^\dagger &
      a^T\end{array}\right]P
\left[\begin{array}{c}a \\ a^\#\end{array}\right],\frac{1}{2}\left[\begin{array}{cc}a^\dagger &
      a^T\end{array}\right]M
\left[\begin{array}{c}a \\ a^\#\end{array}\right]\right]} \nonumber \\
&=& \left[\begin{array}{c}a \\ a^\#\end{array}\right]^\dagger 
\left[
PJM - MJP 
\right] \left[\begin{array}{c}a \\ a^\#\end{array}\right].
\end{eqnarray*}
Also,
\begin{eqnarray*}
\lefteqn{ \frac{1}{2}L^\dagger[\left[\begin{array}{cc}a^\dagger &
      a^T\end{array}\right]P
\left[\begin{array}{c}a \\ a^\#\end{array}\right],L]}\nonumber \\
\lefteqn{+\frac{1}{2}[L^\dagger,\left[\begin{array}{cc}a^\dagger &
      a^T\end{array}\right]P
\left[\begin{array}{c}a \\ a^\#\end{array}\right]]L} \nonumber \\
&=& \tr\left(PJN^\dagger\left[\begin{array}{cc}I & 0 \\ 0 & 0 \end{array}\right]NJ\right)
\nonumber \\&&
-\frac{1}{2}\left[\begin{array}{c}a \\ a^\#\end{array}\right]^\dagger
\left(N^\dagger J N JP+PJN^\dagger J N\right)
\left[\begin{array}{c}a \\ a^\#\end{array}\right].
\end{eqnarray*}
\end{lemma}
{\em Proof:}
The proof of these identities follows via  straightforward but tedious
calculations using (\ref{CCR2}). \hfill $\Box$

\begin{lemma}
\label{L6}
Suppose $z$ is defined as in (\ref{z}) and $L$ is defined as in (\ref{L}).  Then for any positive definite matrix $P$ of the form (\ref{Pform}) and any Hermitian matrix 
$M$ of the form (\ref{Mform}),
\begin{eqnarray*}
\lefteqn{-i[z,\frac{1}{2}\left[\begin{array}{cc}a^\dagger &
      a^T\end{array}\right]M
\left[\begin{array}{c}a \\ a^\#\end{array}\right]]}\nonumber \\
\lefteqn{+\frac{1}{2}\left(L^\dagger[z,L] + [L^\dagger,z]L\right)}\nonumber \\ &=& \tilde E\left(-iJM-\frac{1}{2} JN^\dagger JN \right)
\left[\begin{array}{c}a \\ a^\#\end{array}\right]\\
&=& \tilde E F \left[\begin{array}{c}a \\ a^\#\end{array}\right]
\end{eqnarray*}
where 
\begin{equation}
\label{F}
F = -iJM-\frac{1}{2} JN^\dagger JN. 
\end{equation}
Furthermore, 
\[
i[z,\left[\begin{array}{cc}a^\dagger &
      a^T\end{array}\right]P
\left[\begin{array}{c}a \\ a^\#\end{array}\right]] = 2i\tilde E JP\left[\begin{array}{c}a \\ a^\#\end{array}\right].
\]
\end{lemma}
{\em Proof:}
The proof of these equations  follows via  straightforward but tedious
calculations using (\ref{CCR2}). \hfill $\Box$

\begin{lemma}
\label{L7}
Given a complex row vector $\tilde T= [T_1~T_2]$. Then
\[
\tilde T\left[\begin{array}{c}a \\ a^\#\end{array}\right] = \left[\begin{array}{c}a \\ a^\#\end{array}\right]^\dagger \Sigma \tilde T^T.
\] 
\end{lemma}
{\em Proof:}
The proof of this result  follows via  straightforward 
calculations. \hfill $\Box$

\noindent
{\em Proof of Theorem \ref{T2}.}
If the conditions of the theorem are satisfied, then the transfer function
$\frac{\gamma}{2} - (1+\theta s)G(s)$ is strictly positive real. However, this transfer function has a state space realization
\[
\frac{\gamma}{2} - (1+\theta s)G(s) \sim 
\left[\begin{array}{c|c} F & G\\
\hline \\
-H - \theta H F & \frac{\gamma}{2} -\theta H G 
\end{array}\right]
\]
where $F = -iJM-\frac{1}{2} JN^\dagger JN$, $G = 2iJ \Sigma \tilde E^T$ and $H = \tilde E^\# \Sigma$. It now follows using the strict positive real lemma that the linear matrix inequality
\begin{equation}
\label{SPRLMI}
\left[\begin{array}{cc} 
PF + F^\dagger P & P G + H^\dagger + \theta F^\dagger H^\dagger \\
G^\dagger P + H+\theta H F & -\gamma + \theta(HG+G^\dagger H^\dagger)\end{array} \right] < 0
\end{equation}
will have a solution $P > 0$ of the form (\ref{Pform}).  This matrix $P$ defines a corresponding Lyapunov operator $V \in \mathcal{P}$ as in (\ref{quadV}). Furthermore, it is straightforward to verify that $HG+G^\dagger H^\dagger = 0$. Hence, using Schur complements, it follows from (\ref{SPRLMI}) that
\begin{eqnarray}
\label{SPRQMI}
\lefteqn{PF + F^\dagger P}\nonumber \\
&&+ \frac{1}{\gamma}\left( P G + H^\dagger + \theta F^\dagger H^\dagger\right)
\left(G^\dagger P + H+\theta H F\right) < 0.\nonumber \\
\end{eqnarray}

Now using Lemma \ref{L6} and Lemma \ref{L7}, we have
\begin{eqnarray*}
\lefteqn{i[z,\left[\begin{array}{cc}a^\dagger &
      a^T\end{array}\right]\left(P-\frac{\theta}{2} M\right)
\left[\begin{array}{c}a \\ a^\#\end{array}\right]]}\nonumber \\
\lefteqn{+\frac{\theta}{2}\left(L^\dagger[z,L] + [L^\dagger,z]L\right)+z}\\
 &=& i[z,\left[\begin{array}{cc}a^\dagger &
      a^T\end{array}\right]P
\left[\begin{array}{c}a \\ a^\#\end{array}\right]] \nonumber \\
&&+\theta\left(\begin{array}{c}-i[z,\frac{1}{2}\left[\begin{array}{cc}a^\dagger &
      a^T\end{array}\right]M
\left[\begin{array}{c}a \\ a^\#\end{array}\right]]\\
+\frac{1}{2}\left(L^\dagger[z,L] + [L^\dagger,z]L\right)\end{array} \right)+z\\
&=& \left(2i\tilde EJP +\theta \tilde E F   +\tilde E \right)\left[\begin{array}{c}a \\ a^\#\end{array}\right]\\
&=& \left[\begin{array}{c}a \\ a^\#\end{array}\right]^\dagger \Sigma \left(2i\tilde EJP +\theta \tilde E F   +\tilde E \right)^T\\
&=& \left[\begin{array}{c}a \\ a^\#\end{array}\right]^\dagger 
\left(2i PJ +\theta F^\dagger   + I \right)\Sigma \tilde E^T.
\end{eqnarray*}
Hence using Lemma \ref{L2}, we obtain
\begin{eqnarray}
\label{lyap_ineq3}
\lefteqn{-i[\left[\begin{array}{cc}a^\dagger &
      a^T\end{array}\right]P
\left[\begin{array}{c}a \\ a^\#\end{array}\right],\frac{1}{2}\left[\begin{array}{cc}a^\dagger &
      a^T\end{array}\right] M
\left[\begin{array}{c}a \\ a^\#\end{array}\right]]}\nonumber \\&&
+ \frac{1}{2}L^\dagger[\left[\begin{array}{cc}a^\dagger &
      a^T\end{array}\right]P
\left[\begin{array}{c}a \\ a^\#\end{array}\right],L]\nonumber \\&&
+\frac{1}{2}[L^\dagger,\left[\begin{array}{cc}a^\dagger &
      a^T\end{array}\right]P
\left[\begin{array}{c}a \\ a^\#\end{array}\right]]L\nonumber \\ &&
+\frac{1}{\gamma}\left(\begin{array}{c}i[z,\left[\begin{array}{cc}a^\dagger &
      a^T\end{array}\right]\left(P-\frac{\theta}{2} M\right)
\left[\begin{array}{c}a \\ a^\#\end{array}\right]] \\
+\frac{\theta}{2}\left(L^\dagger[z,L] + [L^\dagger,z]L\right)+z\end{array}\right)
\nonumber \\&&\times
\left(\begin{array}{c}i[z,\left[\begin{array}{cc}a^\dagger &
      a^T\end{array}\right]\left(P-\frac{\theta}{2} M\right)
\left[\begin{array}{c}a \\ a^\#\end{array}\right]] \\
+\frac{\theta}{2}\left(L^\dagger[z,L] + [L^\dagger,z]L\right)+z\end{array}\right)^*\nonumber\\
&=& \left[\begin{array}{c}a \\ a^\#\end{array}\right]^\dagger\tilde M \left[\begin{array}{c}a \\
a^\#\end{array}\right]\nonumber \\
&&+\tr\left(PJN^\dagger\left[\begin{array}{cc}I & 0 \\ 0 & 0 \end{array}\right]NJ\right)
\end{eqnarray}
where 
\begin{eqnarray*}
\lefteqn{\tilde M=}\\
&&PF + F^\dagger P+ \\
&&\frac{1}{\gamma}\left(2i PJ +\theta F^\dagger   + I \right)\Sigma \tilde E^T
\tilde E^\# \Sigma \left(-2i JP +\theta F  + I \right)\\
&=&PF + F^\dagger P\nonumber \\
&&+ \frac{1}{\gamma}\left( P G + H^\dagger + \theta F^\dagger H^\dagger\right)
\left(G^\dagger P + H+\theta H F\right),
\end{eqnarray*}
$F = -iJM - \frac{1}{2}JN^\dagger J N$, $G = 2iJ \Sigma \tilde E^T$ and $H = \tilde E^\# \Sigma$.
Also, it follows from (\ref{SPRQMI}) that $\tilde M < 0$. 

We now write $V \in \mathcal{P}$ as
\[
V = \bar V + \theta f(z,z^*)
\]
where
\[
\bar V = \left[\begin{array}{cc}a^\dagger &
      a^T\end{array}\right]P
\left[\begin{array}{c}a \\ a^\#\end{array}\right]. 
\]
Also, we define 
\[
\bar H = \frac{1}{2}\left[\begin{array}{cc}a^\dagger &
      a^T\end{array}\right] M
\left[\begin{array}{c}a \\ a^\#\end{array}\right].
\]
 Hence (\ref{lyap_ineq3}) can be re-written as
\begin{eqnarray}
\label{lyap_ineq3a}
\lefteqn{-i[\bar V,\bar H]
+ \frac{1}{2}L^\dagger[\bar V,L]
+\frac{1}{2}[L^\dagger,\bar V]L}\nonumber \\ &&
+\frac{1}{\gamma}\left(\begin{array}{c}i[z,\bar V - \theta \bar H] \\
+\frac{\theta}{2}\left(L^\dagger[z,L] + [L^\dagger,z]L\right)+z\end{array}\right)
\nonumber \\&&\times
\left(\begin{array}{c}i[z,\bar V - \theta \bar H] \\
+\frac{\theta}{2}\left(L^\dagger[z,L] + [L^\dagger,z]L\right)+z\end{array}\right)^*\nonumber\\
&=& \left[\begin{array}{c}a \\ a^\#\end{array}\right]^\dagger\tilde M \left[\begin{array}{c}a \\
a^\#\end{array}\right]\nonumber \\
&&+\tr\left(PJN^\dagger\left[\begin{array}{cc}I & 0 \\ 0 & 0 \end{array}\right]NJ\right).
\end{eqnarray}
Now, it follows from Lemma \ref{LB} and Lemma \ref{LB1} that
\begin{eqnarray}
\label{lyap_ineq3b}
\lefteqn{-\imath[V,H] + \frac{1}{2}L^\dagger[V,L]+\frac{1}{2}[L^\dagger,V]L} \nonumber \\
&=& 
-i[\bar V+\theta f(z,z^*),\bar H+f(z,z^*)] \nonumber \\
&&+ \frac{1}{2}L^\dagger[\bar V+\theta f(z,z^*),L]+\frac{1}{2}[L^\dagger,\bar V+\theta f(z,z^*)]L
\nonumber \\
&=&
-i[\bar V,\bar H]-i\theta[f(z,z^*),\bar H]-i[\bar V,f(z,z^*)]\nonumber \\
&&-i\theta[f(z,z^*),f(z,z^*)]+\frac{1}{2}L^\dagger[\bar V,L]+\frac{1}{2}[L^\dagger,\bar V]L \nonumber \\&&
+ \frac{\theta}{2}L^\dagger[f(z,z^*),L]+\frac{\theta}{2}[L^\dagger,f(z,z^*)]L \nonumber \\
&=& -i[\bar V,\bar H]-i[\bar V-\theta \bar H,f(z,z^*)] \nonumber \\&&
+\frac{1}{2}L^\dagger[\bar V,L]+\frac{1}{2}[L^\dagger,\bar V]L \nonumber \\&&
+ \frac{\theta}{2}L^\dagger[f(z,z^*),L]+\frac{\theta}{2}[L^\dagger,f(z,z^*)]L\nonumber \\
&=& -i[\bar V,\bar H]+\frac{1}{2}L^\dagger[\bar V,L]+\frac{1}{2}[L^\dagger,\bar V]L\nonumber \\
&&-i[\bar V-\theta \bar H,z]w_1+iw_1^*[z^*,\bar V-\theta \bar H]
\nonumber \\&&
-\frac{i}{2}\left[z,[\bar V-\theta \bar H,z]\right]w_2
+\frac{i}{2}w_2^*\left[z,[\bar V-\theta \bar H,z]\right]^*\nonumber \\
&&+\frac{\theta}{2}\left(L^\dagger[z,L] + [L^\dagger,z]L\right)w_1\nonumber \\
&&+\frac{\theta}{2}w_1^* \left(L^\dagger[z^*,L] + [L^\dagger,z^*]L\right) \nonumber \\
&& -\frac{\theta}{2}\left[[L^\dagger,z]L,z\right]w_2\nonumber \\
&&-\frac{\theta}{2}w_2^*\left[[L^\dagger,z]L,z\right]^*\nonumber \\
&& +\frac{\theta}{2}\left[z^*,[L^\dagger,z]L\right]w_3\nonumber \\
&&+\frac{\theta}{2}w_3^*\left[z^*,[L^\dagger,z]L\right]^*\nonumber \\
&=& -i[\bar V,\bar H]+\frac{1}{2}L^\dagger[\bar V,L]+\frac{1}{2}[L^\dagger,\bar V]L\nonumber \\
&& + \left(i[z,\bar V-\theta \bar H]+\frac{\theta}{2}\left(L^\dagger[z,L] + [L^\dagger,z]L\right) \right)w_1\nonumber \\
&& + w_1^*\left(i[z^*,\bar V-\theta \bar H]+\frac{\theta}{2} \left(L^\dagger[z^*,L] + [L^\dagger,z^*]L\right) \right)\nonumber \\
&& -\frac{1}{2}\left(\begin{array}{c}i\left[z,[\bar V-\theta \bar H,z]\right]\\
+ \theta\left[[L^\dagger,z]L,z\right]\end{array}\right)w_2\nonumber \\
&&-\frac{1}{2}w_2^*\left(\begin{array}{c}-i\left[z,[\bar V-\theta \bar H,z]\right]^*\\
+ \theta\left[[L^\dagger,z]L,z\right]^*\end{array}\right)\nonumber \\
&& +\frac{1}{2}\theta\left[z^*,[L^\dagger,z]L\right]w_3\nonumber \\
&&+\frac{1}{2}\theta w_3^*\left[z^*,[L^\dagger,z]L\right]^*.
\end{eqnarray}

Also, $$[\bar V-\theta \bar H,z]^* = z^*(\bar V-\theta \bar H)-(\bar V-\theta \bar H)z^*=[z^*,\bar V-\theta \bar H]$$ since $\bar V-\theta \bar H$ is self-adjoint. 
Therefore, 
\begin{eqnarray*}
0 &\leq& \left(\begin{array}{c} -\frac{i[z,\bar V-\theta \bar H]+\frac{\theta}{2}\left(L^\dagger[z,L] + [L^\dagger,z]L\right)}{\sqrt{\gamma}}\\
+  \sqrt{\gamma}\left(w_1-z^*/\gamma\right)^*\end{array}\right)\\
 &&\times\left(\begin{array}{c} -\frac{i[z,\bar V-\theta \bar H]+\frac{\theta}{2}\left(L^\dagger[z,L] + [L^\dagger,z]L\right)}{\sqrt{\gamma}}\\
+  \sqrt{\gamma}\left(w_1-z^*/\gamma\right)^*
\end{array}\right)^*\nonumber \\
&=&  \frac{1}{\gamma}\left(i[z,\bar V-\theta \bar H]+\frac{\theta}{2}\left(L^\dagger[z,L] + [L^\dagger,z]L\right)\right) \nonumber \\
&&\times\left(i[z^*,\bar V-\theta \bar H]+\frac{\theta}{2} \left(L^\dagger[z^*,L] + [L^\dagger,z^*]L\right)\right)\\
&&-\left(\begin{array}{c}i[z,\bar V-\theta \bar H]\\+\frac{\theta}{2}\left(L^\dagger[z,L] + [L^\dagger,z]L\right)\end{array}\right)\left(w_1-z^*/\gamma\right)\\
&&-\left(w_1-z^*/\gamma\right)^*\left(\begin{array}{c}i[z^*,\bar V-\theta \bar H]\\+\frac{\theta}{2} \left(L^\dagger[z^*,L] + [L^\dagger,z^*]L\right) \end{array}\right)\\
&&+\gamma\left(w_1-z^*/\gamma\right)^*\left(w_1-z^*/\gamma\right)
\end{eqnarray*}
and hence
\begin{eqnarray}
\label{ineq3a}
\lefteqn{\left(i[z,\bar V-\theta \bar H]+\frac{\theta}{2}\left(L^\dagger[z,L] + [L^\dagger,z]L\right)\right)w_1}\nonumber \\
\lefteqn{+w_1^*\left(i[z^*,\bar V-\theta \bar H]+\frac{\theta}{2} \left(L^\dagger[z^*,L] + [L^\dagger,z^*]L\right)\right)}\nonumber \\
& \leq & \frac{1}{\gamma}\left(i[z,\bar V-\theta \bar H]+\frac{\theta}{2}\left(L^\dagger[z,L] + [L^\dagger,z]L\right)\right)\nonumber \\
&&\times\left(i[z^*,\bar V-\theta \bar H]+\frac{\theta}{2} \left(L^\dagger[z^*,L] + [L^\dagger,z^*]L\right)\right)\nonumber\\
&&+\gamma\left(w_1-z^*/\gamma\right)^*\left(w_1-z^*/\gamma\right)
\nonumber\\
&&\left(i[z,\bar V-\theta \bar H]+\frac{\theta}{2}\left(L^\dagger[z,L] + [L^\dagger,z]L\right)\right)z^*/\gamma\nonumber \\
&&+z\left(i[z^*,\bar V-\theta \bar H]+\frac{\theta}{2} \left(L^\dagger[z^*,L] + [L^\dagger,z^*]L\right)\right)/\gamma\nonumber \\
&=& \frac{1}{\gamma}\left(i[z,\bar V-\theta \bar H]+\frac{\theta}{2}\left(L^\dagger[z,L] + [L^\dagger,z]L\right)+z\right)
\nonumber \\
&&\times\left(i[z^*,\bar V-\theta \bar H]+\frac{\theta}{2} \left(L^\dagger[z^*,L] + [L^\dagger,z^*]L\right)+z^*\right)\nonumber\\
&&+\gamma\left(w_1-z^*/\gamma\right)^*\left(w_1-z^*/\gamma\right)
-\frac{zz^*}{\gamma}.
\end{eqnarray}
Furthermore, 
\begin{eqnarray*}
0 &\leq& 
\left(\left(\theta\left[[L^\dagger,z]L,z\right]+i\left[z,[\bar V-\theta \bar H,z]\right]\right)^*+ w_2\right)^* \nonumber \\
&& \times \left(\left(\theta\left[[L^\dagger,z]L,z\right]+i\left[z,[\bar V-\theta \bar H,z]\right]\right)^*+ w_2\right)\nonumber \\
&=&  \left(\theta\left[[L^\dagger,z]L,z\right]+i\left[z,[\bar V-\theta \bar H,z]\right]\right) \nonumber \\
&& \times \left(\theta\left[[L^\dagger,z]L,z\right]+i\left[z,[\bar V-\theta \bar H,z]\right]\right)^*\nonumber \\
&&
+ w_2^*\left(\theta\left[[L^\dagger,z]L,z\right]+i\left[z,[\bar V-\theta \bar H,z]\right]\right)^*\nonumber \\
&&+\left(\theta\left[[L^\dagger,z]L,z\right]+i\left[z,[\bar V-\theta \bar H,z]\right]\right)w_2+ w_2^* w_2
\end{eqnarray*}
and hence
\begin{eqnarray}
\label{ineq3b}
\lefteqn{-\left(\theta\left[[L^\dagger,z]L,z\right]+i\left[z,[\bar V-\theta \bar H,z]\right]\right)w_2}\nonumber \\
\lefteqn{-w_2^*\left(\theta\left[[L^\dagger,z]L,z\right]^*-i\left[z,[\bar V-\theta \bar H,z]\right]^*\right)}\nonumber\\
&\leq& \left(\theta\left[[L^\dagger,z]L,z\right]+i\left[z,[\bar V-\theta \bar H,z]\right]\right) \nonumber \\
&&\times \left(\theta\left[[L^\dagger,z]L,z\right]+i\left[z,[\bar V-\theta \bar H,z]\right]\right)^*\nonumber \\
&&+ w_2^* w_2.
\end{eqnarray}
Similarly
\begin{eqnarray*}
0 &\leq& 
\left(\theta\left[z^*,[L^\dagger,z]L\right]^*- w_3\right)^*\nonumber \\
&&\times \left(\theta\left[z^*,[L^\dagger,z]L\right]^*- w_3\right)\nonumber \\
&=&  \theta^2\left[z^*,[L^\dagger,z]L\right]\left[z^*,[L^\dagger,z]L\right]^*\nonumber \\
&& - w_3^*\theta\left[z^*,[L^\dagger,z]L\right]^*
-\theta\left[z^*,[L^\dagger,z]L\right]w_3 + w_3^* w_3
\end{eqnarray*}
and hence
\begin{eqnarray}
\label{ineq3c}
\lefteqn{\theta\left[z^*,[L^\dagger,z]L\right]w_3+w_3^*\theta\left[z^*,[L^\dagger,z]L\right]^*}\nonumber \\
&\leq&\theta^2\left[z^*,[L^\dagger,z]L\right]\left[z^*,[L^\dagger,z]L\right]^* + w_3^* w_3.
\end{eqnarray}

Substituting (\ref{ineq3a}), (\ref{ineq3b}) and (\ref{ineq3c}) into (\ref{lyap_ineq3b}), it follows that
\begin{eqnarray}
\label{ineq2a}
\lefteqn{-\imath[V,H] + \frac{1}{2}L^\dagger[V,L]+\frac{1}{2}[L^\dagger,V]L} \nonumber \\
 &\leq &  
-i[\bar V,\bar H]+\frac{1}{2}L^\dagger[\bar V,L]+\frac{1}{2}[L^\dagger,\bar V]L\nonumber \\
&&+\frac{1}{\gamma}\left(i[z,\bar V-\theta \bar H]+\frac{\theta}{2}\left(L^\dagger[z,L] + [L^\dagger,z]L\right)+z\right)
\nonumber \\
&&\times\left(i[z^*,\bar V-\theta \bar H]+\frac{\theta}{2} \left(L^\dagger[z^*,L] + [L^\dagger,z^*]L\right)+z^*\right)\nonumber\\
&&+\gamma\left(w_1-z^*/\gamma\right)^*\left(w_1-z^*/\gamma\right)
-\frac{zz^*}{\gamma}\nonumber \\
&&+\frac{1}{2}\left(\theta\left[[L^\dagger,z]L,z\right]+i\left[z,[\bar V-\theta \bar H,z]\right]\right) \nonumber \\
&&\times \left(\theta\left[[L^\dagger,z]L,z\right]+\frac{1}{2}i\left[z,[\bar V-\theta \bar H,z]\right]\right)^*\nonumber \\
&&+ \frac{1}{2}w_2^* w_2
\nonumber \\
&&+\frac{\theta^2}{2}\left[z^*,[L^\dagger,z]L\right]\left[z^*,[L^\dagger,z]L\right]^* + \frac{1}{2}w_3^* w_3
\nonumber \\
&\leq&
-i[\bar V,\bar H]+\frac{1}{2}L^\dagger[\bar V,L]+\frac{1}{2}[L^\dagger,\bar V]L\nonumber \\
&&+\frac{1}{\gamma}\left(i[z,\bar V-\theta \bar H]+\frac{\theta}{2}\left(L^\dagger[z,L] + [L^\dagger,z]L\right)+z\right)
\nonumber \\
&&\times\left(i[z^*,\bar V-\theta \bar H]+\frac{\theta}{2} \left(L^\dagger[z^*,L] + [L^\dagger,z^*]L\right)+z^*\right)\nonumber\\
&&+\frac{\theta^2}{2}\left[z^*,[L^\dagger,z]L\right]\left[z^*,[L^\dagger,z]L\right]^*\nonumber \\
&&+\frac{1}{2}\left(\theta\left[[L^\dagger,z]L,z\right]+i\left[z,[\bar V-\theta \bar H,z]\right]\right) \nonumber \\
&& \times \left(\theta\left[[L^\dagger,z]L,z\right]+i\left[z,[\bar V-\theta \bar H,z]\right]\right)^*\nonumber \\&&
+\delta_1\gamma+\frac{\delta_2}{2}+\frac{\delta_3}{2} 
\end{eqnarray}
using (\ref{sector2}), (\ref{smooth1}), (\ref{smooth2}), and (\ref{w1w2w3}).
 Then it follows from (\ref{lyap_ineq3a}) that 
\begin{eqnarray*}
\lefteqn{-\imath[V,H] + \frac{1}{2}L^\dagger[V,L]+\frac{1}{2}[L^\dagger,V]L} \nonumber \\
&\leq&\left[\begin{array}{c}a \\ a^\#\end{array}\right]^\dagger\tilde M \left[\begin{array}{c}a \\
a^\#\end{array}\right]\nonumber \\
&&+\tr\left(PJN^\dagger\left[\begin{array}{cc}I & 0 \\ 0 & 0 \end{array}\right]NJ\right)\nonumber \\
&&+\frac{\theta^2}{2}\left[z^*,[L^\dagger,z]L\right]\left[z^*,[L^\dagger,z]L\right]^*\nonumber \\
&&+\frac{1}{2}\left(\theta\left[[L^\dagger,z]L,z\right]+i\left[z,[\bar V-\theta \bar H,z]\right]\right) \nonumber \\
&& \times \left(\theta\left[[L^\dagger,z]L,z\right]+i\left[z,[\bar V-\theta \bar H,z]\right]\right)^*\nonumber \\&&
+\delta_1\gamma+\frac{\delta_2}{2}+\frac{\delta_3}{2}.
\end{eqnarray*}
Since $\tilde M > 0$, it follows using  (\ref{upper_lower}) that there exists a constant $c > 0$ such that 
\begin{eqnarray*}
\lefteqn{-\imath[V,H] + \frac{1}{2}L^\dagger[V,L]+\frac{1}{2}[L^\dagger,V]L+cV} \nonumber \\
&\leq& -\imath[V,H] + \frac{1}{2}L^\dagger[V,L]+\frac{1}{2}[L^\dagger,V] + c\left(\bar V +    \theta\beta z^*z\right)\\
& \leq & \tr\left(PJN^\dagger\left[\begin{array}{cc}I & 0 \\ 0 & 0 \end{array}\right]NJ\right)\nonumber \\
&&+\frac{\theta^2}{2}\left[z^*,[L^\dagger,z]L\right]\left[z^*,[L^\dagger,z]L\right]^*\nonumber \\
&&+\frac{1}{2}\left(\theta\left[[L^\dagger,z]L,z\right]+i\left[z,[\bar V-\theta \bar H,z]\right]\right) \nonumber \\
&& \times \left(\theta\left[[L^\dagger,z]L,z\right]+i\left[z,[\bar V-\theta \bar H,z]\right]\right)^*\nonumber \\&&
+\delta_1\gamma+\frac{\delta_2}{2}+\frac{\delta_3}{2}.
\end{eqnarray*}
That is,
\[
-\imath[V,H] + \frac{1}{2}L^\dagger[V,L]+\frac{1}{2}[L^\dagger,V]L+cV \leq \lambda
\]
where
\begin{eqnarray*}
 \lambda &=& \tr\left(PJN^\dagger\left[\begin{array}{cc}I & 0 \\ 0 & 0 \end{array}\right]NJ\right)\\
&&+\frac{\theta^2}{2}\left[z^*,[L^\dagger,z]L\right]\left[z^*,[L^\dagger,z]L\right]^*\nonumber \\
&&+\frac{1}{2}\left(\theta\left[[L^\dagger,z]L,z\right]+i\left[z,[\bar V-\theta \bar H,z]\right]\right) \nonumber \\
&& \times \left(\theta\left[[L^\dagger,z]L,z\right]+i\left[z,[\bar V-\theta \bar H,z]\right]\right)^*\nonumber \\
&&+\delta_1\gamma+\frac{\delta_2}{2}+\frac{\delta_3}{2}\nonumber \\
&=&\tr\left(PJN^\dagger\left[\begin{array}{cc}I & 0 \\ 0 & 0 \end{array}\right]NJ\right)\\
&&+\frac{\theta^2}{2}\tilde E^\# J \Sigma\tilde N^\dagger \tilde N^\# \Sigma J \tilde E^\dagger\tilde E J \Sigma\tilde N^T \tilde N \Sigma J \tilde E^T\\
&&+\frac{1}{2}\left(\begin{array}{c}-\theta\tilde E J \tilde N^\dagger \tilde N \Sigma J \tilde E^T\\
+i\tilde E \Sigma J\left(P-\frac{\theta}{2} M\right)J\tilde E^T \end{array} \right) \nonumber \\
&&\times \left(\begin{array}{c}-\theta\tilde E J \tilde N^\dagger \tilde N \Sigma J \tilde E^T\\
+i\tilde E \Sigma J\left(P-\frac{\theta}{2} M\right)J\tilde E^T \end{array} \right)^* \nonumber \\
&&+\delta_1\gamma+\frac{\delta_2}{2}+\frac{\delta_3}{2}\nonumber \\
&\geq & 0
\end{eqnarray*}
using Lemma \ref{L4} and Lemma \ref{L5}. Therefore, it follows from  Lemma \ref{L0},  and  $P > 0$ that 
\begin{eqnarray}
\label{ms_stable1}
\lefteqn{\left< \left[\begin{array}{c}a(t) \\ a^\#(t)\end{array}\right]^\dagger \left[\begin{array}{c}a(t) \\ a^\#(t)\end{array}\right] \right> \leq}\nonumber \\
&&  e^{-ct}\left< \left[\begin{array}{c}a(0) \\ a^\#(0)\end{array}\right]^\dagger \left[\begin{array}{c}a(0) \\ a^\#(0)\end{array}\right] \right>\frac{\lambda_{max}[P+\beta \tilde E^\dagger \tilde E]}{\lambda_{min}[P]}\nonumber \\
&&+ \frac{\lambda}{c\lambda_{min}[P]}~~\forall t \geq 0.
\end{eqnarray} 
 Hence, the condition (\ref{ms_stable0}) is satisfied with $c_1 = \frac{\lambda_{max}[P+\beta \tilde E^\dagger \tilde E]}{\lambda_{min}[P]} > 0$, $c_2 = c > 0$ and $c_3 = \frac{\lambda}{c\lambda_{min}[P]} \geq 0$. 
\hfill $\Box$

\begin{observation}
\label{O1}
Note that  
the SPR condition  (\ref{SPR}) can be re-written as
\begin{eqnarray}
\label{SPR2a}
\frac{\gamma}{2} +\mathcal{R}e[G(i\omega)] -\theta \omega\mathcal{I}m[G(i\omega)]  &>& 0
\end{eqnarray}
for all $ \omega \in [-\infty,\infty]$. The condition (\ref{SPR2a}),
 can be tested graphically by producing a plot of $\omega
\mathcal{I}m[G(i\omega)]$ versus $\mathcal{R}e[G(i\omega)]$ with
$\omega\in [-\infty,\infty]$ as a parameter. Such a parametric plot
is referred to as the  Popov plot; e.g., see \cite{KHA02}. Then, the condition (\ref{SPR2a}),
 will be satisfied if and only if the Popov plot lies below the
 straight line of slope $\frac{1}{\theta}$ and with $x$-axis intercepts
$-\frac{\gamma}{2}$; see Figure \ref{F2}. 

\begin{figure}[htbp]
\begin{center}
\psfrag{Im}{$\omega\mathcal{I}m[G(i\omega)]$}
\psfrag{Re}{$ \mathcal{R}e[G(i\omega)]$}
\psfrag{slope}{slope $= \frac{1}{\theta}$}
\psfrag{g4}{$\frac{\gamma}{2}$}
\psfrag{mg4}{$-\frac{\gamma}{2}$}
\psfrag{ar}{allowable region}
\includegraphics[width=6cm]{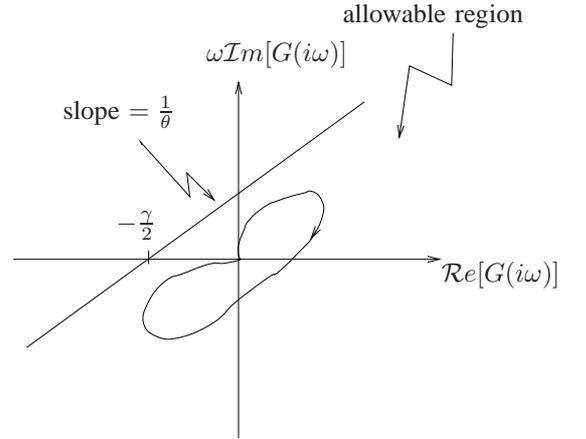}
\end{center}
\caption{Allowable region for the Popov plot.}
\label{F2}
\end{figure}
\end{observation}

\section{Analysis of an optical cavity containing a Kerr medium}
\label{sec:example}
\begin{figure}[hbp]
\begin{center}
\includegraphics[width=8cm]{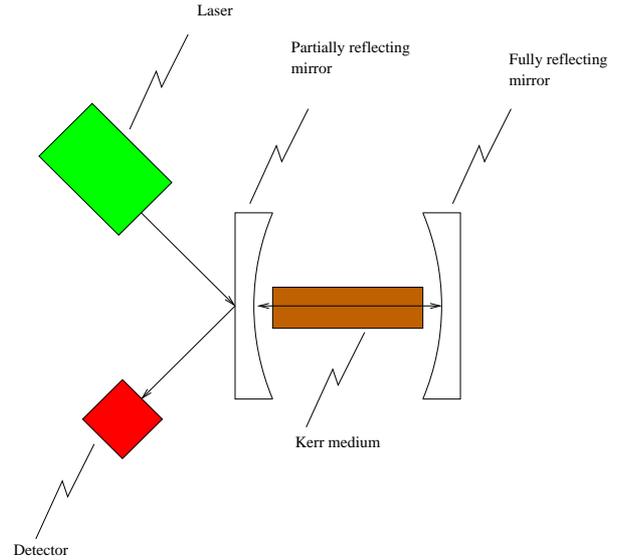}
\end{center}
\caption{Schematic diagram of an optical cavity containing a Kerr medium.}
\label{F1}
\end{figure}

The system under consideration consists of an optical cavity containing a Kerr medium. The optical cavity is made from  two mirrors, one of which is partially reflecting and one of which is fully reflecting. The cavity is driven by a laser beam directed at the partially reflecting mirror. The corresponding reflected beam is then measured using a detector. The Kerr medium within the cavity can be constructed from a suitable nonlinear optical crystal; e.g., see \cite{BR04}. This system is illustrated in Figure \ref{F1}. 

A standard $(S,L,H)$ model for an optical cavity containing a Kerr medium  is as follows:
\begin{equation}
\label{Kerr}
S=I,~H = \left(a^*\right)^2a^2,~L
= \sqrt{\kappa}a;
\end{equation}
e.g., see \cite{WM08}.  We first attempt to apply the results of Theorem \ref{T1} and Theorem \ref{T2} to  this quantum system. Hence, we let
\[
M = 0
\]
and
\[
f(z,z^*)= z^2 \left(z^*\right)^2
\]
where $z = a^*$. 
This defines a nonlinear quantum
system of the form considered in Theorem \ref{T1} and Theorem \ref{T2} with $M_1 = 0$, $M_2 =
0$, $N_1 = \sqrt{\kappa}$, $N_2 = 0$, $E_1=0$, $E_2 = 1$.  We now investigate whether this function $f(\cdot)$ satisfies the conditions 
(\ref{upper_lower}), (\ref{sector1}), (\ref{sector2}), (\ref{smooth1}), and (\ref{smooth2}). Now, 
\begin{eqnarray*}
\frac{\partial f(z,z^*)}{\partial z} &=& 2z\left(z^*\right)^2;\\
\frac{\partial^2f(z,z^*)}{\partial z^2} &=& 2\left(z^*\right)^2;\\
\frac{\partial^2f(z,z^*)}{\partial z\partial z ^*} &=& 4zz^*.
\end{eqnarray*}
Also, the sector condition (\ref{sector2}) can be rewritten as 
\begin{eqnarray*}
\lefteqn{\gamma\frac{\partial f(z,z^*)}{\partial z}^*\frac{\partial f(z,z^*)}{\partial z}}\\
 &\leq&\frac{\partial f(z,z^*)}{\partial z}^*z^*+z\frac{\partial f(z,z^*)}{\partial z} +\gamma\delta_1
\end{eqnarray*}
which is satisfied for $\gamma = 0$ since 
\[
\frac{\partial f(z,z^*)}{\partial z}^*z^*+z\frac{\partial f(z,z^*)}{\partial z} = 4z^2\left(z^*\right)^2 \geq 0.
\]
However this condition is not satisfied for any finite value of $\gamma > 0$. Also,  conditions (\ref{sector1}), (\ref{upper_lower}), (\ref{sector2}),  (\ref{smooth1}), (\ref{smooth2}) are not satisfied. 

In order to overcome this difficulty, we note that any physical realization of a Kerr nonlinearity will not be exactly described by the model (\ref{Kerr}) but rather will exhibit some saturation of the Kerr effect; e.g., see \cite{BBBDS12}. In order to represent this effect, we will assume that the true  function $ \tilde f(\cdot)$ describing the Hamiltonian of the Kerr medium  is such that its Taylor series expansion (\ref{H2nonquad}) satisfies $S_{0,k} = S_{1,k} = 0$ for all $k= 0,1, \ldots$ and $S_{2,2} = 1$. That is, the first non-zero term in the Taylor series expansion corresponds to the standard Kerr Hamiltonian given in (\ref{Kerr}). Furthermore, we assume that the function $\tilde f(\cdot)$ is such that the conditions 
(\ref{sector1}), (\ref{sector2}),  (\ref{smooth1}), (\ref{smooth2}), (\ref{upper_lower})  are all satisfied for suitable values of the constants 
 $\gamma > 0$, $\beta > 0$, $\delta_1\geq 0$, $\delta_2\geq 0$, $\delta_3\geq 0$. Here the quantity $\frac{1}{\gamma}$ will be proportional to the saturation limit for the Kerr nonlinearity.  Thus, under these assumptions, we can assume  $\tilde f(\cdot)  \in \mathcal{W}_1$ and $\tilde f(\cdot)  \in \mathcal{W}_2$. 

\begin{figure}[htbp]
\begin{center}
\includegraphics[width=8cm]{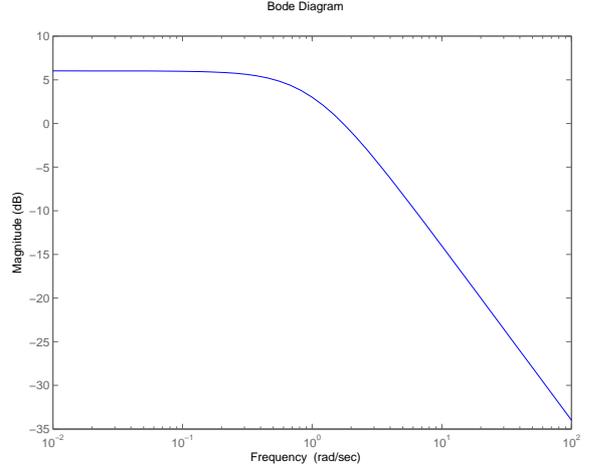}
\end{center}
\caption{Magnitude Bode plot of $G(s)$ for the case $\kappa =2$.}
\label{F4}
\end{figure}

This system has $F =\left[\begin{array}{cc}
  -\frac{\kappa}{2} & 0 \\0 & -\frac{\kappa}{2}\end{array}\right]$, which is Hurwitz for all $\kappa > 0$ and  $ G(s) = -\frac{2i}{s +\frac{\kappa}{2}}$. A magnitude Bode plot of this transfer function, is shown in Figure \ref{F4} for the case of $\kappa =2$. 
In this case, we obtain $\|G(s)\|_\infty = 2$ and in general
\[
\|G(s)\|_\infty = \frac{4}{\kappa}.
\]
Thus, applying Theorem \ref{T1} to this system, we can guarantee that the system is  mean
square stable provided
\begin{equation}
\label{cond1}
\kappa > \frac{4}{\gamma}.
\end{equation}

\begin{figure}[htbp]
\begin{center}
\psfrag{Im}{\tiny $\omega\mathcal{I}m[G(i\omega)]$}
\psfrag{Re}{\tiny $ \mathcal{R}e[G(i\omega)]$}
\includegraphics[width=8cm]{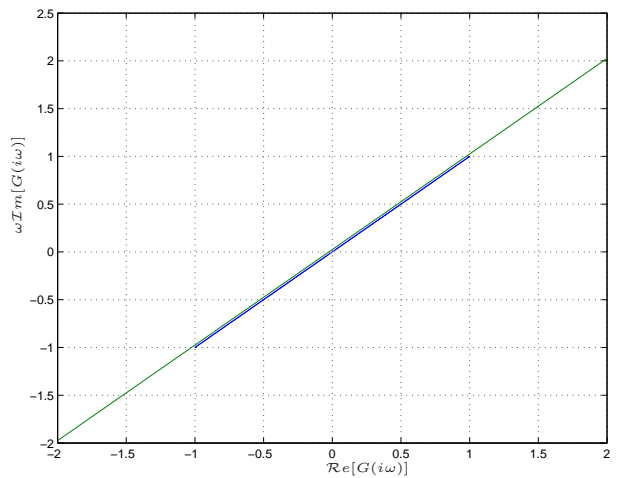}
\end{center}
\caption{Popov plot for the Kerr nonlinearity system with $\kappa = 2$ and $\gamma = 0.1$.}
\label{F3}
\end{figure}

\begin{figure}[htbp]
\begin{center}
\psfrag{Im}{\tiny $\omega\mathcal{I}m[G(i\omega)]$}
\psfrag{Re}{\tiny $ \mathcal{R}e[G(i\omega)]$}
\includegraphics[width=8cm]{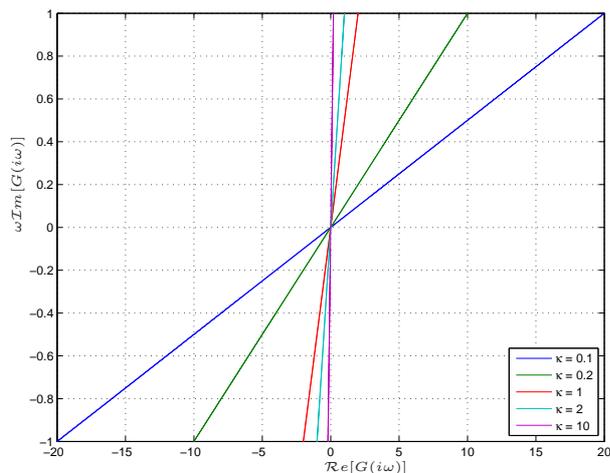}
\end{center}
\caption{Popov plot for the Kerr nonlinearity system with different values of $\kappa > 0$.}
\label{F5}
\end{figure}

We now apply our new result Theorem \ref{T2} to further analyze the stability of the system. We first choose $\kappa =
  2$ and construct the  Popov plot corresponding to the
  transfer function $G(s)$ as
  discussed in Observation \ref{O1}. For a value of $\theta = 1$,
  this plot, along with the corresponding allowable region corresponding to $\gamma = 0.1$, is shown in
  Figure \ref{F3}. From this figure it can be seen that the 
  Popov plot lies in the allowable region and hence, it follows
  from Theorem \ref{T2} and Observation \ref{O1} that this system will be mean
square stable for $\kappa = 2$ and $\gamma = 0.1$. In fact, it follows from this plot that the frequency domain condition (\ref{SPR}) will be satisfied for all $\gamma > 0$. This condition is clearly less restrictive than the condition (\ref{cond1}) obtained by applying Theorem \ref{T1}. Furthermore, we can construct the Popov plot of the system for different values of $\kappa > 0$ as shown in Figure \ref{F5}. From these plots, we can see that for a suitable value of $\theta = \frac{2}{\kappa} > 0$, the frequency domain condition (\ref{SPR}) will be satisfied for all $\gamma > 0$ and all $\kappa > 0$. Thus, using Theorem \ref{T2}, we can conclude that the optical cavity containing a saturated Kerr medium is in fact mean square stable for all $\gamma > 0$ and $\kappa > 0$. 

\section{Conclusions}
\label{sec:conclusions}
In this paper, we have introduced a new nonlinear quantum Popov stability criterion and applied it to the robust stability
analysis of a nonlinear quantum system consisting of an optical cavity containing a Kerr medium. We have also applied an existing quantum small gain theorem to the analysis of this system. By choosing a model which represents a saturating Kerr medium, both approaches to robust stability analysis were applicable to this system. Furthermore both approaches were able to verify the robust mean square stability of this system for some range of parameter values. However, the quantum small gain theorem approach was found to be more conservative than the quantum Popov criterion approach in that it could only verify robust mean square stability for a restricted range of parameters. In contrast, the quantum Popov approach was able to verify the robust mean square stability of the system for all positive values of the system parameters.

\bibliography{/home/irp/Bibliog/irpnew}
\bibliographystyle{IEEEtran}

\end{document}